\documentclass[aps,prl,10pt,twocolumn,longbibliography,noeprint,floatfix,superscriptaddress]{revtex4-2}

\usepackage{amsmath, amssymb}
\usepackage{mathtools}
\usepackage{lipsum}
\usepackage{times}
\usepackage{graphicx}
\usepackage{wasysym}
\usepackage{bm}
\usepackage{hyperref}
\usepackage{xspace}
\usepackage{siunitx}
\usepackage[dvipsnames]{xcolor}
\usepackage{soul}
\usepackage{centernot}
\usepackage{braket}
\usepackage{microtype}
\usepackage{upgreek}
\usepackage{booktabs}
\usepackage{float}

\definecolor{customcolor}{rgb}{0.261, 0.212, 0.658}

\definecolor{myColor2}{rgb}{0.02,0.12,0.3}  

\definecolor{myColor}{rgb}{0.02,0.12,0.7}
\definecolor{myciteColor}{rgb}{0.39,0.7,0.89}
\hypersetup{colorlinks=true, linkcolor=myColor, filecolor=myColor, urlcolor=myColor, citecolor=myColor, urlcolor=myColor}
\DeclareSIUnit{\gauss}{\ensuremath{\mathrm{G}}}
\DeclareSIUnit{\bohrradius}{\ensuremath{a_0}}
\graphicspath{{./Figures/}}

\DeclareSIUnit{\nK}{\nano\kelvin}
\DeclareSIUnit{\um}{\micro\metre}
\DeclareSIUnit{\aB}{\emph{a}_0}
\DeclareSIUnit{\G}{G}

\newcommand{\appropto}{\mathrel{\vcenter{
  \offinterlineskip\halign{\hfil$##$\cr
    \propto\cr\noalign{\kern2pt}\sim\cr\noalign{\kern-2pt}}}}}

\def\be{\begin{equation}}
\def\ee{\end{equation}}

\makeatletter
\def\@fnsymbol#1{\ensuremath{\ifcase#1\or *\or \dagger\or \ddagger\or
   \mathsection\or \mathparagraph\or \|\or **\or \dagger\dagger
   \or \ddagger\ddagger \else\@ctrerr\fi}}
\makeatother

\newcommand{\um}{\upmu{\rm m}}
\newcommand{\nK}{\textrm{nK}}
\newcommand{\kB}{k_{\textrm{B}}}

\newcommand{\potassium}{^{39}\textrm{K}}

\begin{document} 
 
\title{
Imaging the vacuum fluctuations of a quantum field
}

\author{Yansheng~Zhang}\email{yz661@cam.ac.uk}
\affiliation{Cavendish Laboratory, University of Cambridge, J. J. Thomson Avenue, Cambridge CB3 0US, UK}
\author{Feiyang~Wang}
\affiliation{Cavendish Laboratory, University of Cambridge, J. J. Thomson Avenue, Cambridge CB3 0US, UK}
\author{{Yi~Jiang}}
\affiliation{Cavendish Laboratory, University of Cambridge, J. J. Thomson Avenue, Cambridge CB3 0US, UK}
\author{Alexander~C.~Jenkins}
\affiliation{Kavli Institute for Cosmology, University of Cambridge, Madingley Road, Cambridge CB3 0HA, UK}
\affiliation{DAMTP, University of Cambridge, Wilberforce Road, Cambridge CB3 0WA, UK}
\author{Paul~H.~C.~Wong}
\affiliation{Cavendish Laboratory, University of Cambridge, J. J. Thomson Avenue, Cambridge CB3 0US, UK}
\author{Christoph~Eigen}
\affiliation{Cavendish Laboratory, University of Cambridge, J. J. Thomson Avenue, Cambridge CB3 0US, UK}
\author{{Gehrig~Carlse}} 
\affiliation{Cavendish Laboratory, University of Cambridge, J. J. Thomson Avenue, Cambridge CB3 0US, UK}
\author{Zoran~Hadzibabic}
\affiliation{Cavendish Laboratory, University of Cambridge, J. J. Thomson Avenue, Cambridge CB3 0US, UK}

\begin{abstract}
Heisenberg uncertainties lead to inevitable fluctuations in the measurement outcomes for quantum-mechanical observables. 
For quantum fields, these uncertainties result in random spatial structures in snapshots of a field, even when the field is in its ground (vacuum) state.
Such `vacuum fluctuations' are at the heart of a wide range of phenomena, from spontaneous decay processes to the Casimir force and Hawking radiation. 
Their existence is a key manifestation of the quantumness of the physical world, but usually it is only their consequences that are directly observed. 
Here, we directly observe spatial vacuum fluctuations of a bosonic quantum field. 
Our experiments are based on a homogeneous planar atomic Bose--Einstein condensate. The condensate comprises two coherently coupled interacting components (spin states), and the quantum field describes its spin degrees of freedom.
In the regime where the interactions dominate over the coherent coupling, our system emulates a (massive relativistic) sine-Gordon field.
Images of the field reveal simultaneous fluctuations on different length scales, with scale-dependent amplitudes consistent with theoretical predictions for a vacuum state.
Observing such fluctuations in the sine-Gordon limit opens many possibilities for laboratory simulations of relativistic fields in regimes that are presently not theoretically tractable. 
\end{abstract}
\maketitle

\begin{figure*}
    \centering
    \includegraphics[width=\linewidth]{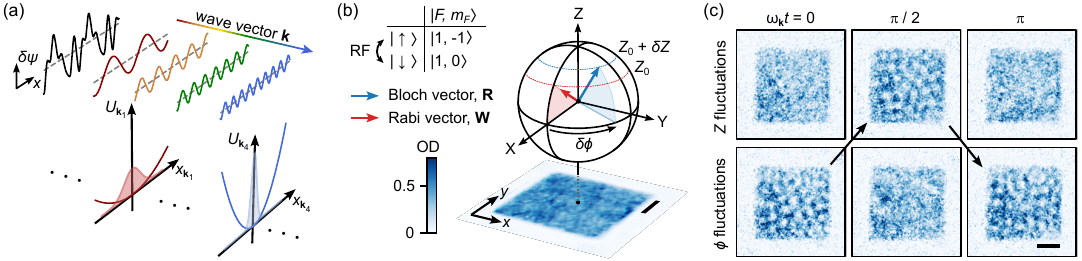}
    \caption{
    \textbf{Quantum-field fluctuations.}   
    (a)~The concept. In a uniform system, fluctuations of a quantum field $\psi$ (black), here illustrated in one spatial dimension, can  be decomposed into plane-wave normal modes (colors) with wave vectors $\mathbf{k}$. Each such mode is equivalent to a quantum harmonic oscillator with a $\mathbf{k}$-dependent frequency $\omega_\mathbf{k}$, (generalized) displacement $x_\mathbf{k}$ and momentum $p_\mathbf{k}$, and a potential $U_\mathbf{k}\propto \omega_\mathbf{k}^2 x_\mathbf{k}^2$.
    Even in its ground state, each oscillator exhibits quantum uncertainties; the color shading indicates Gaussian $x_\mathbf{k}$ uncertainties $\propto 1/\sqrt{\omega_\mathbf{k}}$. 
    (b)~The experiment. 
    We encode a massive bosonic field in the spin degrees of freedom of a homogeneous planar condensate (bottom image; OD is the optical density), produced in an optical box trap. Our condensate comprises two hyperfine states of $^{39}$K atoms, denoted $\ket{\uparrow}$ and $\ket{\downarrow}$, coherently coupled by a radio-frequency (RF) field. The coupling strength $\Omega$ and RF detuning $\delta$ define the spatially constant Rabi vector ${\bf W}$. The total atom density is essentially uniform, while the spatially varying spin state is represented by local Bloch vectors ${\bf R}(x,y)$, defined by the population imbalance $Z(x,y)$ and relative phase $\phi(x,y)$ of the two components. Classically, the lowest energy state, $\delta Z (x,y) = \delta \phi (x,y) = 0$, corresponds to a constant ${\bf R}(x,y)$ aligned with ${\bf W}$.
(c)~Observing $Z$ and $\phi$ fluctuations. Here we parametrically excite modes with $|\mathbf{k}| \approx 1.3\,\mathrm{\um}^{-1}$ and $\omega_\mathbf{k} \approx 2\pi\times2~\mathrm{kHz}$. The magnitudes of $Z$ and $\phi$ fluctuations, revealed by in-situ state-selective imaging, oscillate in quadrature at $2\omega_\mathbf{k}$. Each image is taken in a separate experimental run. The scale bars in (b) and (c) correspond to $10\,\um$.
}
\label{fig:1}
\end{figure*}

Quantum field theories underpin our understanding of the inherently quantum physical world, from quantum optics~\cite{Loudon:1983} and condensed matter physics~\cite{Fradkin:2013} to the phenomenology of elementary particles~\cite{Peskin:1995} and models of inflationary cosmology~\cite{Mukhanov:2005}.
A defining feature of quantum fields is the existence of their inherent fluctuations. 
Each spatial mode of the field can be mapped onto a quantum harmonic oscillator [see Fig.~\ref{fig:1}(a)], described by a pair of conjugate variables that obey Heisenberg-like uncertainty relations. Under repeated measurements, or more generally in interactions with another physical system, observables associated with the quantum field can appear to fluctuate even if the field is in its ground (vacuum) state.
These so-called vacuum fluctuations have profound physical consequences. They are responsible for the spontaneous decay of excited atomic states~\cite{CohenTannoudji:1992b}, the Lamb shift of atomic energy levels~\cite{Lamb:1947}, and the Casimir force between mesoscopic objects~\cite{Casimir:1948}. They also underlie the theory of Hawking radiation from black holes~\cite{Hawking:1974}, and are believed to have seeded, prior to cosmic inflation, the large-scale inhomogeneities seen in the present-day universe~\cite{Springel:2006}. However, directly observing vacuum fluctuations, rather than just their consequences, remains a formidable challenge. 

For electromagnetic fields, evidence of vacuum fluctuations was observed via electro-optic sampling~\cite{Riek:2015, Benea-Chelmus:2019}. For atomic matter fields, the possibility of in-situ imaging enables more detailed access to field fluctuations. This allowed, for example, studies of their signatures in the local counting statistics~\cite{Chuu:2005, Jacqmin:2011, Armijo:2012,Xiang:2025,Yao:2025,deJongh:2025}, their effect on spatial mode entanglement~\cite{Fadel:2018, Kunkel:2018, Lange:2018, Chen:2021, TiZhang:2024}, and their response to external modulations or changes in inter-atomic interactions~\cite{Jaskula:2012, Hung:2013, Clark:2017, ZZhang:2020, Chen:2021, Viermann:2022, Sparn:2024, Liebster:2025a, Liebster:2025b, Tamura:2026}.

Here, we directly image the random spatial structures arising from scale-dependent fluctuations of a massive bosonic field that describes the spin degrees of freedom of a planar two-component Bose--Einstein condensate (BEC), and show that their spectrum agrees with the expected vacuum fluctuations.

Our experiment [see Fig.~\ref{fig:1}(b)] is based on a homogeneous miscible mixture of two hyperfine states (labeled $\ket{\uparrow}$  and $\ket{\downarrow}$) of $\potassium$ atoms in a two-dimensional (2D) optical box trap~\cite{Chomaz:2015, Navon:2021, Christodoulou:2021}, as described in~\cite{YZhang:2026}. We work in the deeply degenerate regime, 
where our system is an essentially pure finite-size condensate, described by a two-component wave function $\psi =(\psi_\uparrow, \psi_\downarrow)^T$, with $\psi_{\uparrow,\downarrow} = \sqrt{n(1\pm Z)/2}~\exp[-i(\theta\pm \phi/2)]$. 
Here $(\theta, n)$ and $(\phi, Z)$ are pairs of conjugate variables describing, respectively, the density and spin degrees of freedom; $\theta(x, y)$ is the common-mode phase of the two components, $n(x, y)$ their total density, $\phi(x, y)$ their relative phase, and $Z(x, y)$ their population imbalance.

We focus on the field that describes the spin degrees of freedom, and coherently couple the $\ket{\uparrow}$ and $\ket{\downarrow}$ components using a radio-frequency (RF) field with Rabi frequency $\Omega$ and detuning $\delta$. In our system, the intra-component interactions differ for $\ket{\uparrow}$ and $\ket{\downarrow}$ states, so the total spin-dependent interaction energy is minimized at a non-zero population imbalance $Z_0 = 0.53$~\cite{YZhang:2026}. 
On the other hand, the Rabi energy is minimized if the Bloch vectors ${\bf R}(x,y)$, describing the local spin states, are all aligned with the Rabi vector $\mathbf{W} = (\Omega, 0, \delta)$. We always set $\delta$ to $\Omega Z_0/\sqrt{1 - Z_0^2}$, so, classically, the state with a spatially uniform $\bf R$ aligned with $\bf W$ minimizes both the interaction and the Rabi energy. However, quantum-mechanically, the ground state includes local deviations, $(\delta\phi, \delta Z)$, between $\bf R$ and $\bf W$.

Around $Z=Z_0$, perturbative spin and density dynamics are decoupled~\cite{Jenkins:2024, YZhang:2026}, and spin fluctuations across different length scales are described by the Lagrangian density,
\begin{equation}
\begin{split}
    2\mathcal{L}/(\alpha \bar{n}) = \: \hbar\,\delta\dot{\phi}~\, \delta Z' - \frac{1}{2}\delta\phi\left[\hbar \Omega' - \frac{\hbar^2\nabla^2}{2m}\right]& \delta\phi~\\
    \quad- \frac{1}{2}\delta Z'\left[(\hbar\Omega' + 2\mu_\text{s}') -\frac{\hbar^2\nabla^2}{2m}\right]& \delta Z'\,,
\end{split}
\label{eq:langrangian}
\end{equation}
where $m$ is the atom mass and $\bar{n}=\langle n\rangle$ is the spatially averaged total atom density. Note that trigonometric factors due to our $Z_0\neq0$ are encapsulated in $\alpha = (1-Z_0^2)=0.72$ and the scaled variables $\delta Z'=\delta Z/\alpha$, $\Omega'=  \Omega/\sqrt{\alpha}$, and $\mu_\text{s}'=\alpha\mu_\text{s}$, where $\mu_\text{s}\simeq h\times900~$Hz characterizes the spin-dependent interaction energy.

For our spatially uniform system, field fluctuations are naturally decomposed into plane-wave modes with wave vectors $\mathbf{k}$, which evolve independently according to
\begin{equation}
\begin{split}
    \hbar\delta\dot{\phi}_\mathbf{k} &= (\epsilon_\mathbf{k}+\hbar\Omega' +  2\mu_{\text{s}}') \delta Z_\mathbf{k}'\,,\\
    \hbar\delta\dot{Z}_\mathbf{k}' &=- (\epsilon_\mathbf{k}  + \hbar\Omega') \delta \phi_\mathbf{k}\,,
\end{split}
\label{eq:EoM fourier}
\end{equation}
where $\epsilon_{\mathbf{k}} = \hbar^2|\mathbf {k}|^2/(2m)$ and ($\delta\phi_\mathbf{k}$, $\delta Z_\mathbf{k}'$) are the Fourier components of ($\delta\phi$, $\delta Z'$)~\cite{fourierFootnote}.
Each $\mathbf{k}$ mode is formally equivalent to a harmonic oscillator with a generalized displacement $x_\mathbf{k} \propto \delta\phi_\mathbf{k}$ and momentum $p_\mathbf{k} \propto\delta Z'_\mathbf{k}$.
The corresponding mode frequency gives the dispersion relation of the field
\begin{equation}
    \hbar\omega_\mathbf{k} = \sqrt{(\epsilon_\mathbf{k} + \hbar\Omega'  + 2\mu_{\text{s}}')(\epsilon_\mathbf{k} + \hbar\Omega' )}\,,
\label{eq:dispersion}
\end{equation}
which depends only on $k=|\mathbf{k}|$. In our experiment, we tune the dispersion by varying $\Omega'$, and we focus on the long wavelength limit $\epsilon_\mathbf{k}<\mu_\text{s}'$. For $\hbar\Omega'/\mu_\text{s}'\gg 1$, the field is in the Rabi regime, with a nearly flat dispersion $\omega_\mathbf{k}\simeq\Omega'$, while for $\hbar\Omega'/\mu_\text{s}'\ll 1$, the field is in the Josephson regime and has a massive relativistic dispersion~\cite{Cominotti:2022, YZhang:2026}.
In the latter case, in the non-perturbative regime beyond small displacements, the $\phi(x,y)$ field experiences a $\cos{\phi}$ potential and encodes the sine-Gordon model~\cite{Son:2002,YZhang:2026}, allowing analog simulation of many phenomena in cosmology~\cite{Kofman:1994, Vilenkin:2000, Marsh:2016, Chatrchyan:2021} and high-energy physics~\cite{Weinberg:2012}. 

So far we have treated $\delta\phi_\mathbf{k}$ and $\delta Z_\mathbf{k}'$ as classical variables, but they are actually non-commuting operators that describe the underlying quantum field, and their canonical conjugation relation, $[\delta\phi_\mathbf{k}^\dagger,\delta Z'_{\mathbf{k}'}]=2i\delta_{\mathbf{k}\mathbf{k}'}/(\alpha \bar{n})$, is analogous to $[x_\mathbf{k},p_{\mathbf{k}'}]=i\hbar\delta_{\mathbf{k}\mathbf{k}'}$ for an ensemble of independent quantum harmonic oscillators.
Consequently,  even when the field is in its ground state, each of its modes exhibits Heisenberg uncertainties
\begin{equation}
\begin{split}
    \langle|\delta\phi_\mathbf{k}|^2\rangle &= \frac{1}{\alpha \bar{n}}\sqrt{\frac{\epsilon_{\mathbf k} + \hbar\Omega'+2\mu_{\text{s}}'}{\epsilon_{\mathbf k} + \hbar\Omega'}}\,, \\
    \langle|\delta Z_\mathbf{k}'|^2\rangle &= \frac{1}{\alpha \bar{n}}\sqrt{\frac{\epsilon_{\mathbf k} + \hbar\Omega'}{\epsilon_{\mathbf k} + \hbar\Omega'+2\mu_{\text{s}}'}}\,.
\end{split}
\label{eq:vacuum fluctuations}
\end{equation}
Note that the corresponding real-space fluctuations, such as  $\langle\delta \phi^2\rangle = \int \langle|\delta\phi_\mathbf{k}|^2\rangle \text{d}^2\mathbf{k}/(2\pi)^2$, scale as $\bar{n}^{-1}$, and $\bar{n}\xrightarrow{}\infty$ corresponds to the classical limit.
\begin{figure*}[t]
    \centering
    \includegraphics[width=\linewidth]{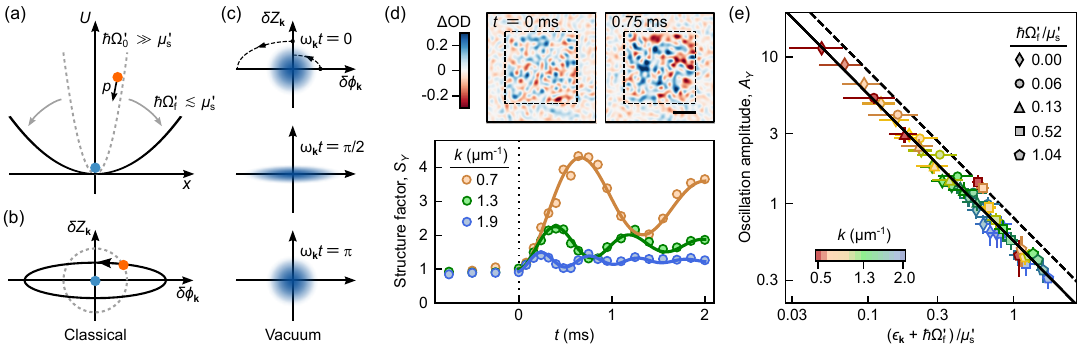}
    \caption{
\textbf{Revealing the effect of vacuum fluctuations in the Rabi regime.}
(a-c)~Experimental concepts.
(a)~We prepare the system deeply in the Rabi regime ($\Omega' = \Omega_0'\gg\mu_{\text{s}}'/\hbar$), and at $t=0$ quench $\Omega'$  to $\Omega_{\text{f}}'\lesssim \mu_{\text{s}}'/\hbar$, effectively decompressing the harmonic oscillator potentials associated with each spin mode. 
(b)~Classically, the quench increases the ellipticity of the phase-space orbit of a particle at finite energy (orange dot), but leaves a zero-energy particle unaffected (blue dot).
(c)~Quantum mechanically, the ground state of a spin mode is represented by a distribution of points in phase space (blue shading). Following the quench, each point evolves along its elliptical classical trajectory, which results in $\langle|\delta \phi_{\bf{k}}|^2\rangle$ and $\langle|\delta Z'_{\bf{k}}|^2\rangle$ oscillating at frequency $2\omega_\mathbf{k}$.
(d)~Example measurements, for $\hbar\Omega_{\rm f}'/\mu'_{\rm s}=0.06$. At different times $t$, we image $n_{-i} = |\psi_\uparrow+i\psi_\downarrow|^2/2$, which reveals $\phi$ fluctuations. The top panels show example images before and after the quench; here the mean of approximately $70$ images has been subtracted, and, to suppress photon shot noise, we apply a low-pass filter that retains only modes with $k<2~\um^{-1}$. The dashed lines indicate the boundaries of the box trap; the scale bar corresponds to $10\,\um$. For each $t$, from approximately $70$ images we extract the structure factor $S_Y(k, t)$, which measures $\langle|\delta \phi_{\bf{k}}|^2\rangle$ (see text). 
The bottom panel shows oscillations of $S_Y(k, t)$ for different $k$. The solid lines show damped-sinusoidal fits, with theoretically expected $\omega_\mathbf{k}$, from which we extract the oscillation amplitudes $A_Y(k)$.
(e)~Log-log plot of the extracted $A_Y$ versus $(\epsilon_\mathbf{k} + \hbar\Omega'_{\rm f})/\mu'_{\rm s}$, for different $k$ (colors) and $\Omega'_{\rm f}$ (symbols). The dashed line shows the theoretical prediction for oscillations seeded by vacuum fluctuations in the initial state (see text), and the data follow the same trend (solid line). The systematic difference between the data and the theory (a factor of $\eta \approx 0.7$) likely arises due to imperfect atom detection, which reduces the observed two-particle correlations~\cite{corroFootnote}.
}
\label{fig:2}
\end{figure*}

We probe spin fluctuations by in-situ state-selective imaging. Measurements of the densities $n_{\uparrow,\downarrow} = |\psi_{\uparrow, \downarrow}|^2$ directly reveal $\delta Z$. To measure $\delta\phi$, just before imaging we apply RF pulses that uniformly rotate the Bloch vectors $\mathbf{R}(x,y)$, interfering the two components~\cite{Sadler:2006}. 
In Fig.~\ref{fig:1}(c), we show images that reveal $\delta Z(x,y)$ and $\delta \phi(x,y)$ for a system where modes with a specific $k \simeq \SI{1.3}{\per \micro \meter}$ are parametrically excited; here, for the $\delta\phi$ images, all Bloch vectors are rapidly rotated by $\pi/2$ around the Rabi vector $\mathbf{W}$ just before imaging the $\ket{\downarrow}$ component.
As the system evolves, $Z$ and $\phi$ fluctuations oscillate in quadrature, just like $p$ and $x$ in a classical harmonic oscillator.

\begin{figure*}[t]
    \centering
    \includegraphics[width=\linewidth]{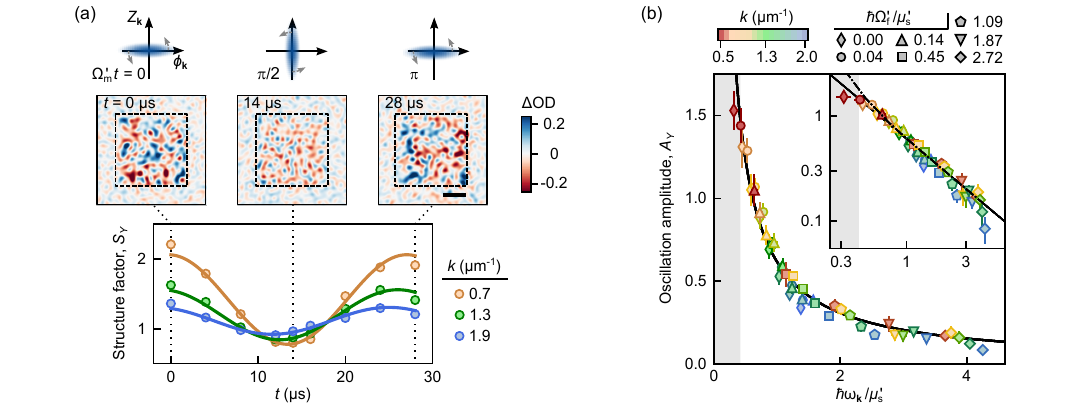}
    \caption{
\textbf{Imaging vacuum fluctuations.}
(a)~Experimental protocol and example measurements. Starting from the Rabi regime, we adiabatically ramp down the coupling strength to $\Omega'_\text{f}$.
For $\hbar\Omega'_\text{f}/\mu_\text{s}'\lesssim 1$, the ground state of each spin mode has an elliptical distribution in phase space.
At $t = 0$, we jump up the coupling strength to $\hbar\Omega_\text{m}'/\mu_\text{s}'\simeq20$, initiating a rotation of the phase-space distribution at $\Omega_\text{m}'$ (top panels). 
The middle panels show example images of $n_{-i}$ for $\hbar\Omega_\text{f}'/\mu_\text{s}'=0.04$, taken as the distribution rotates; here, as in Fig.~\ref{fig:2}(d), we subtract the mean image and low-pass filter the images, and the dashed lines indicate the boundaries of the box trap; the scale bar corresponds to $10\,\um$.
The bottom panel shows oscillations of $S_Y(k, t)$ for different $k$, extracted from approximately $120$ images for each $t$.
The solid lines show sinusoidal fits from which we extract the oscillation amplitudes $A_Y(k)$.
(b)~Spin fluctuation spectrum. The extracted $A_Y(k)$ show a clear $\propto \omega_\mathbf{k}^{-1}$ scaling, as expected for vacuum field fluctuations. The solid line shows the theoretical prediction (see text) incorporating $\eta$, as determined from Fig.~\ref{fig:2}(e). The shaded region indicates the range where our adiabaticity criterion is not satisfied (see text). The inset shows the same data plotted on a log-log scale. A fit (outside the shaded region) allowing for a non-zero temperature (dot-dashed line) gives $T=(6\pm6)$\,nK.
}
    \label{fig:3}
\end{figure*}

For quantitative studies of spin fluctuations, we first focus on the Rabi regime (Fig.~\ref{fig:2}).
Here we initialize the spin field in its ground state deeply in the Rabi regime, with $\Omega'=\Omega_{0}' \simeq 10\mu_{\text{s}}'/\hbar$. To prepare this state, we start with a pure $\ket{\uparrow}$ state and adiabatically rotate $\mathbf{R}$ to the target population imbalance $Z = Z_0$, using a $3$-ms partial Landau--Zener transfer.

In this regime, the expected vacuum fluctuations, $ \langle|\delta\phi_\mathbf{k}|^2\rangle\simeq \langle|{\delta}Z'_\mathbf{k}|^2\rangle\simeq(\alpha \bar{n})^{-1}$, are $k$-independent and insensitive to the exact value of the large $\Omega'$; see Eq.~(\ref{eq:vacuum fluctuations}). To reveal these fluctuations, we quench $\Omega'$ to various values $\Omega_\text{f}'\lesssim \mu_\text{s}'/\hbar$, which results in a $k$-dependent amplification of $\phi$ fluctuations and suppression of the $Z$ ones; see Fig.~\ref{fig:2}(a--c).

The quench, at $t=0$, effectively decompresses the harmonic oscillator potentials associated with each spin mode~[Fig.~\ref{fig:2}(a)]. Classically~[Fig.~\ref{fig:2}(b)], this does not affect a zero-energy `particle' (blue), while a finite energy one (orange) hops from a circular phase-space trajectory to an elliptical one, with the ratio of the $\delta\phi_\mathbf{k}$ and $\delta Z'_\mathbf{k}$ oscillation amplitudes given by $r_\mathbf{k}=\sqrt{(\epsilon_\mathbf{k}+\hbar\Omega'+2\mu_\text{s}')/(\epsilon_\mathbf{k}+\hbar\Omega')}$; see Eq.~(\ref{eq:EoM fourier}).
Quantum mechanically~[Fig.~\ref{fig:2}(c)], even the ground state has a finite spread in phase space, and for $t>0$, each point in this distribution follows its post-quench classical trajectory. At $\omega_\mathbf{k}t = \pi/2$, the initial ${\delta}Z_\mathbf{k}'$ is rotated into ${\delta}\phi_\mathbf{k}$ with an amplification factor  $r_\mathbf{k}$, and the initial ${\delta}\phi_\mathbf{k}$ is rotated into ${\delta}Z_\mathbf{k}'$ with a suppression factor $1/r_\mathbf{k}$. At $\omega_\mathbf{k}t=\pi$, the distribution returns to its initial state, and more generally the fluctuation strengths oscillate at frequency $2\omega_\mathbf{k}$.
The peak-to-peak amplitude of $\langle|\delta\phi_\mathbf{k}|^2\rangle$ oscillations seeded by vacuum fluctuations is
\begin{equation}
    A_\phi(\mathbf{k})\,= \frac{r_\mathbf{k}^2-1}{\alpha \bar{n}} = \frac{1}{\alpha \bar{n}} \cdot \frac{2\mu_{\text{s}}'}{\epsilon_\mathbf{k} + \hbar\Omega_\text{f}'}\, .
    \label{eq:QuenchDown}
\end{equation}

Experimentally, at different times after the quench, we perform a rapid $\pi/2$ spin rotation around the $X$ axis of the Bloch sphere and then image the $\ket{\uparrow}$ component. This implements a projective measurement along the $Y$ axis of the Bloch sphere, which gives $n_{-i} = |\psi_\uparrow+i\psi_\downarrow|^2/2$. The spatial variations in $n_{-i}$ provide the maximal signal for fluctuations in $\sin\phi\simeq \phi$.

In Fig.~\ref{fig:2}(d), we show examples of our measurements, for ${\hbar}\Omega_\text{f}'/\mu_\text{s}'\simeq0.06$. The images of $n_{-i}$ clearly show the enhancement of $\phi$ fluctuations. These fluctuations are quantified by the structure factor
\begin{equation}
\begin{split}
    S_Y(\mathbf{k},t) & = \int \frac{\langle n_{-i}(\mathbf{r},{t}) \,n_{-i}(\mathbf{r}',{t}) \rangle}{N_{-i}} e^{-i\mathbf{k}\cdot(\mathbf{r} - \mathbf{r}')} \text{d}^2\mathbf{r} ~\text{d}^2\mathbf{r}' \\
    & = \left[S_\text{d}(\mathbf{k}) + \alpha \bar{n} \langle|\delta\phi_\mathbf{k}(t)|^2\rangle\right]/2 \,.
\end{split}
\label{eq:SY(K)}
\end{equation}
Here, $N_{-i}$ is the total number of atoms detected in the projective measurement, and $S_\text{d}(\mathbf{k})$ is the time-independent contribution to $S_Y(\mathbf{k})$ due to the decoupled fluctuations in the total density. As shown in the bottom panel of Fig.~\ref{fig:2}(d), the azimuthally averaged structure factor, $S_Y(k,t)$, for different $k$, oscillates at $2\omega_\mathbf{k}$, with $k$-dependent (initial) amplitudes $A_Y(k)$.
For oscillations seeded by vacuum fluctuations, one expects 
\begin{equation}
    A_Y(\mathbf{k}) =\alpha \bar{n}A_\phi(\mathbf{k})/2 = \mu_\text{s}'/(\epsilon_\mathbf{k}+\hbar\Omega_\text{f}')\,.
    \label{eq:AYQuenchDown}
\end{equation}

In Fig.~\ref{fig:2}(e), we plot the extracted $A_Y(k)$ for quenches to different $\Omega'_\text{f}$. The dashed line shows the theoretical prediction based on Eq.~(\ref{eq:AYQuenchDown}); here we include a $20\%$ reduction (calculated in a truncated Wigner approximation~\cite{Blakie:2008}) due to the non-infinite values of $\Omega_0'$ in our pre-quench state and $\Omega$ during the $\pi/2$ rotation. 
Our measurements follow the same trend, with the data systematically lower by a factor of $\eta \approx 0.7$, as shown by the solid line. This small difference likely arises due to imperfect atom detection, which reduces the observed two-particle correlations~\cite{corroFootnote}.

In a second set of experiments, we directly observe the vacuum fluctuations without amplifying them (Fig.~\ref{fig:3}).
Here, we first prepare the spin field in its ground state in the Rabi regime ($\hbar\Omega'/\mu_\text{s}'\simeq 2.7$), and then linearly ramp down $\Omega'$ over $9\,$ms to different $\Omega_\text{f}'$. For this slow ramp, $\text{d}\omega_\mathbf{k}/\text{d}t < \omega_\mathbf{k}^2$ for nearly all our data in Fig.~\ref{fig:3}.
This adiabaticity condition ensures that the field modes remain close to their ground states, for which  $ \langle|\delta\phi_\mathbf{k}|^2\rangle $ and $\langle|{\delta}Z'_\mathbf{k}|^2\rangle$ are given by Eq.~(\ref{eq:vacuum fluctuations}).

As shown in Fig.~\ref{fig:3}(a), we image both $ \langle|\delta\phi_\mathbf{k}|^2\rangle $ and $\langle|{\delta}Z'_\mathbf{k}|^2\rangle$ by rotating the Bloch vectors $\mathbf{R}(x, y)$ around the Rabi vector $\mathbf{W}$ before performing the projective measurements as in Fig.~\ref{fig:2}. Here, at $t = 0$, we jump $\Omega'$ to a large $\Omega'_\text{m} \simeq 20\mu_\text{s}'/\hbar$, and map out the oscillation of $S_Y(k,t)$ as the ground-state distribution rotates around $\mathbf W$. 

The oscillation amplitude, $A_Y(\mathbf{k})$, measures the difference $ \langle|\delta\phi_\mathbf{k}|^2\rangle-\langle|{\delta}Z'_\mathbf{k}|^2\rangle$ for the state at $t=0$, and as before, is insensitive to the total density fluctuations. It is given by 
\begin{equation}
    A_Y(\mathbf{k}) = {2\mu_{\text{s}}'}/({\hbar\omega_\mathbf{k}}) \cdot (1/2 + N_\mathbf{k})\, ,
\label{eq:AYRotation}
\end{equation}
where the `quantum $1/2$' in the bracket on the r.h.s.~corresponds to vacuum fluctuations, and follows directly from Eq.~(\ref{eq:vacuum fluctuations}). Here we also allow for possible thermal excitations of the modes; $N_\mathbf{k}$ is the mode occupation number, which in thermal equilibrium at temperature $T$ follows the Bose distribution $N_\mathbf{k}=1/[\exp(\hbar\omega_\mathbf{k}/(\kB T))-1]$.
When vacuum fluctuations dominate ($N_\mathbf{k}\ll 1/2$), the oscillation amplitude exhibits a characteristic scaling $A_Y \propto \omega_\mathbf{k}^{-1}$, while in the thermal limit ($\kB T\gg\hbar\omega_\mathbf{k}$), a large phonon occupation $N_\mathbf{k}\simeq \kB T /(\hbar\omega_\mathbf{k})$ leads to $A_Y \propto \omega_\mathbf{k}^{-2}$ instead.

In Fig.~\ref{fig:3}(b), we show that our $A_Y$, over a wide range of $k$ and $\Omega_{\text{f}}'$, follows the $\omega_{\mathbf{k}}^{-1}$ scaling, as expected for vacuum fluctuations; the shaded region indicates where our adiabaticity condition $\text{d}\omega_\mathbf{k}/\text{d}t < \omega_\mathbf{k}^2$ breaks down. 
The magnitude of $A_Y$ is also consistent with the $T = 0$ theoretical prediction (solid line), assuming the same $\eta$ as in Fig.~\ref{fig:2}(e).
In the inset, we show a fit (for points outside the shaded region) that allows for a non-zero $T$.
This fit essentially overlaps with the zero-temperature prediction, and gives $T=(6\pm6)\,$nK, corresponding to $(0.2 \pm 0.2)\mu_\text{s}'$. 
Note that $\kB T = 0.2\mu_\text{s}'$ corresponds to $N_\mathbf{k} \lesssim 0.1$ for all points outside the shaded region.

Our experiments provide an explicit visual demonstration of the vacuum fluctuations arising from the Heisenberg uncertainty principle, and present an important step towards addressing outstanding problems in quantum field theory through analog simulation.
Vacuum fluctuations underlie the challenges in understanding fundamental processes such as the decay of false vacua~\cite{Coleman:1977a, Callan:1977, Zenesini:2024, Cominotti:2025}, particle production via parametric instabilities~\cite{Yoshimura:1995, Berges:2003}, and defect formation via the quantum Kibble-Zurek mechanism~\cite{Dziarmaga:2010, DelCampo:2014}. Our system provides a platform for exploring such non-perturbative and non-equilibrium phenomena, including cosmological preheating and reheating~\cite{Kofman:1994, Chatrchyan:2021}, and the formation and decay of topological domain-wall networks~\cite{Vilenkin:2000, Weinberg:2012, Marsh:2016}. With the addition of Floquet engineering~\cite{Fialko:2015, Jenkins:2024}, it could also allow simulation of relativistic bubble nucleation in the early universe~\cite{Vilenkin:1983, Guth:1983}. The fact that we can directly observe field fluctuations in the regime where the quantum uncertainty dominates over thermal noise could offer a unique window into the microscopic mechanisms governing these phenomena.

{\bf Acknowledgments}\quad We thank David Tong, Silke Weinfurtner, and Joseph Thywissen for inspiring discussions. Our work was supported by EPSRC [Grant No.~EP/Y01510X/1], ERC [UniFlat], and STFC [Grants No.~ST/T006056/1 and No.~ST/Y004469/1]. 
A.C.J. was supported by EPSRC [Grant No. EP/U536684/1], and by a KICC Gavin Boyle Fellowship.
G.C. was supported by an NSERC Postdoctoral Fellowship.

\section{Supplementary materials}

\subsection{Experimental system}

Our experiments, as described in~\cite{YZhang:2026}, are based on a two-component $^{39}$K BEC. The BEC is harmonically confined (with trapping frequency $\omega_z/(2\pi)\simeq1\,$kHz) to a 2D plane, and trapped in-plane by a square box trap of length $L\approx 33\,\um$ and depth $U_{\rm D}\simeq\kB \times 30$\,nK.
We use the hyperfine states $\ket{\uparrow} = |1, -1\rangle$ and $\ket{\downarrow} = |1, 0\rangle$ in the low-field $|F, m_F\rangle$ basis. We work at a bias magnetic field of $B\simeq58.1~$G, where the scattering lengths $a_{\sigma\sigma'}$, characterizing the contact interaction between atoms in $\sigma, \sigma'\in \{\uparrow, \downarrow\}$, are $a_{\uparrow\uparrow}\simeq31~a_0$, $a_{\uparrow\downarrow}\simeq-53~a_0$, and $a_{\downarrow\downarrow}\simeq220~a_0$~\cite{molscat:2019, Tiemann:2020}; here $a_0$ is the Bohr radius. 

The interaction energy density is
\begin{equation}
    \mathcal{E}_\text{int} = \frac{1}{2}\left(g_{\uparrow\uparrow} n_\uparrow^2 +  2 g_{\uparrow\downarrow} n_\uparrow n_\downarrow + g_{\downarrow\downarrow} n_\downarrow^2 \right)\,,
    \label{eq:int E}
\end{equation}
where $g_{\sigma\sigma'} = (\hbar^2/m)\sqrt{8\pi}\, a_{\sigma\sigma'}/\ell_z$, with $\ell_z=\sqrt{\hbar/(m\omega_z)}$, are the 2D interaction parameters~\cite{Hadzibabic:2011}, and $n_{\uparrow}$ and $n_{\downarrow}$ are the 2D densities of each component.
In terms of $n=n_{\uparrow}+n_{\downarrow}$ and $Z=(n_\uparrow-n_\downarrow)/(n_\uparrow+n_\downarrow)$, Eq.~(\ref{eq:int E}) reads
\begin{equation}
    \mathcal{E}_\text{int}  =\frac{1}{2}g_\text{d} n^2 + \frac{1}{2}(Z - Z_0)^2 \kappa n^2 \, ,
\end{equation}
where $g_\text{d} = (g_{\uparrow\uparrow} g_{\downarrow\downarrow} - g_{\uparrow\downarrow}^2)/(g_{\uparrow\uparrow}+g_{\downarrow\downarrow} - 2g_{\uparrow\downarrow})$, $\kappa=(g_{\uparrow\uparrow}+g_{\downarrow\downarrow} - 2g_{\uparrow\downarrow})/4$, and $Z_0=-\Delta/\kappa$, with $\Delta=(g_{\uparrow\uparrow} - g_{\downarrow\downarrow})/4$. 

For $\Omega=\delta = 0$, the density and spin chemical potentials are, respectively, $\mu_\text{d}=g_\text{d}n$ and $\mu_\text{s}=\kappa n$. For $\Omega,\delta\neq0$, these quantities still provide the relevant interaction energy scales, and their typical values in our experiments are $\mu_\text{d}/h\simeq120~$Hz and $\mu_\text{s}/h\simeq900~$Hz.

For spin-selective imaging, we transfer either the $\ket{\uparrow}$ or $\ket{\downarrow}$ state to the imaging state $\ket{2,-2}$, which has a closed optical transition. We detect atoms in $\ket{2,-2}$ using strong-saturation absorption imaging~\cite{Reinaudi:2007}, which yields the optical density (OD) of the cloud as a measure of the atom density.

\subsection{Extraction of the structure factor}

From a stack of OD images $\rho(x, y)$, measured in independent experimental repetitions, we first compute the real-space variations $\delta\rho(x, y)$ by subtracting the average OD profile, and then calculate the OD noise spectrum using a Fourier transform.
Before taking the Fourier transform, we apply a Gaussian-apodized windowing function~\cite{Gonzalez:2018} to $\delta\rho(x, y)$ to prevent spectral leakage due to the box edges, and use a filtering technique~\cite{Altuntas:2026} based on principal component analysis~\cite{Jolliffe:2002} to suppress fringes in absorption imaging. After the transform, we mask small regions in $\mathbf{k}$-space that are still contaminated by the spurious fringes, and subtract the average photon shot noise, calibrated from OD fluctuations in the $k$-range outside the numerical-aperture limit (N.A.$\,=0.7$) of our microscope~\cite{Hung:2011b}. 

Finally, we azimuthally average the noise spectrum $\mathcal{N}(\mathbf{k})$ to obtain $\mathcal{N}(k)$, and calculate 
\begin{equation}
    S(k) = \mathcal{N}(k) / \mathcal{N}_\text{ref}(k)\,,
    \label{eq:ratio S}
\end{equation}
where $\mathcal{N}_\text{ref}(k)$ is the noise spectrum for a set of reference images of weakly interacting single-component gases that have the same mean OD and fluctuations dominated by uncorrelated atom shot noise.
This normalization absorbs the (weakly non-linear~\cite{Yefsah:2011, Chomaz:2012}) $\rho$ to $n$ conversion factor, and eliminates the effects of microscope aberrations~\cite{Hung:2011b}.

\subsection{Effect of imperfect atom detection efficiency}
The structure factor $S(\mathbf{k})$ can be decomposed as \begin{equation}
    S(\mathbf{k})  = 1 + S_\text{c}(\mathbf{k})\,,
\end{equation}
where the `$1$' arises from uncorrelated shot noise and $S_\text{c}(\mathbf{k})$ is the connected two-particle correlator,
\begin{equation}
    S_\text{c}(\mathbf{k}) = \frac{1}{N}\int\langle\psi^\dagger(\mathbf{r}) \psi^\dagger(\mathbf{r}') \psi(\mathbf{r}') \psi(\mathbf{r})\rangle e^{-i\mathbf{k}\cdot(\mathbf{r}-\mathbf{r}')}\text{d}^2\mathbf{r}\,\text{d}^2\mathbf{r}'\, ,
    \label{eq:Sc}
\end{equation}
where $N$ is the total atom number, and $\psi(\mathbf{r})$ is the atomic field operator~\cite{Pitaevskii:2016}. 

When atoms are detected with an imperfect efficiency $\gamma< 1$, the observed $S_\text{c}(\mathbf{k})$ is attenuated by $\gamma$ because two particles contribute to it only when they are simultaneously detected. Mathematically, the observed atom number is reduced by a factor of $\gamma$ but the integral in Eq.~(\ref{eq:Sc}) is reduced by a factor of $\gamma^2$. 
In contrast, the shot-noise contribution to the observed $S(\mathbf{k})$ remains unity, because imperfect detection introduces additional uncorrelated noise that compensates for the reduced atomic shot-noise when fewer atoms are detected (see~\cite{Steinhauer:2022}). Consequently, the attenuation of the measured $A_Y$ due to $\gamma< 1$ is not removed even after taking the ratio in Eq.~(\ref{eq:ratio S}).

Experimentally, $\gamma< 1$ may originate from imperfect transfer to the imaging state and depumping effects in strong-saturation absorption imaging.


\begin{thebibliography}{70}
\bibitem [{fou()}]{fourierFootnote}%
  \BibitemOpen
  \href@noop {} {}\bibinfo {note} {We define the Fourier component for a
  real-space field variable $O(\mathbf{x})$ as $O_\mathbf{k} = \int
  O(\mathbf{x})e^{-i\mathbf{k}\cdot \mathbf{x}}/L~\text{d}^2\mathbf{x}$, where
  $L^2$ is the area of the system.}\BibitemShut {Stop}%
\end{thebibliography}
\end{document}